\documentclass{article}

\usepackage{microtype}
\usepackage{graphicx}
\usepackage{subfigure}
\usepackage{booktabs} 
\usepackage{svg}
\usepackage{hyperref}

\usepackage{tikz}
\usetikzlibrary{positioning, shapes.geometric, arrows.meta}

\tikzset{
    block/.style={rectangle, draw, fill=blue!20, text width=4.1em, text centered, rounded corners, minimum height=3em, font=\footnotesize},
    line/.style={draw, -Latex, thick}
}

\tikzset{
    output/.style={rectangle, draw, fill=green!20, text width=4.1em, text centered, rounded corners, minimum height=3em, font=\footnotesize},
    line/.style={draw, -Latex, thick}
}

\tikzset{
    loss/.style={rectangle, draw, fill=orange!20, text width=4.1em, text centered, rounded corners, minimum height=3em, font=\footnotesize},
    line/.style={draw, -Latex, thick}
}

\tikzset{
    data/.style={rectangle, draw, text width=4.1em, text centered, rounded corners, minimum height=3em, font=\footnotesize},
    line/.style={draw, -Latex, thick}
}

\usepackage[accepted]{icml2024}

\usepackage{amsmath}
\usepackage{amssymb}
\usepackage{mathtools}
\usepackage{amsthm}

\usepackage[capitalize,noabbrev]{cleveref}

\theoremstyle{plain}

\theoremstyle{definition}

\theoremstyle{remark}

\usepackage[textsize=tiny]{todonotes}

\newcommand{\urlhref}[2]{\texttt{\href{#1}{#2}}}

\icmltitlerunning{A Cross-Modal Approach to Silent Speech with LLM-Enhanced Recognition}

\begin{document}

\twocolumn[
\icmltitle{A Cross-Modal Approach to Silent Speech with LLM-Enhanced Recognition}



\begin{icmlauthorlist}
\icmlauthor{Tyler Benster}{phd}
\icmlauthor{Guy Wilson}{postdoc}
\icmlauthor{Reshef Elisha}{stan}
\icmlauthor{Francis R Willett}{postdoc}
\icmlauthor{Shaul Druckmann}{sch}
\end{icmlauthorlist}

\icmlaffiliation{phd}{Neurosciences PhD Program, Stanford University}
\icmlaffiliation{postdoc}{Department of Neurosurgery, Stanford University}
\icmlaffiliation{stan}{Department of Chemical Engineering, Stanford University}
\icmlaffiliation{sch}{Department of Neurobiology, Stanford University}

\icmlcorrespondingauthor{Tyler Benster}{tbenst@stanford.edu}
\icmlcorrespondingauthor{Shaul Druckmann}{shauld@stanford.edu}

\icmlkeywords{Machine Learning, ICML}

\vskip 0.3in
]




\printAffiliations{}

\begin{abstract}
Silent Speech Interfaces (SSIs) offer a noninvasive alternative to brain-computer interfaces for soundless verbal communication. We introduce Multimodal Orofacial Neural Audio (MONA), a system that leverages cross-modal alignment through novel loss functions---cross-contrast (crossCon) and supervised temporal contrast (supTcon)---to train a multimodal model with a shared latent representation. This architecture enables the use of audio-only datasets like LibriSpeech to improve silent speech recognition. Additionally, our introduction of Large Language Model (LLM) Integrated Scoring Adjustment (LISA) significantly improves recognition accuracy. Together, MONA LISA  reduces the state-of-the-art word error rate (WER) from 28.8\% to 12.2\% in the Gaddy (2020) benchmark dataset for silent speech on an open vocabulary. For vocal EMG recordings, our method improves the state-of-the-art from 23.3\% to 3.7\% WER. In the Brain-to-Text 2024 competition, LISA performs best, improving the top WER from 9.8\% to 8.9\%. To the best of our knowledge, this work represents the first instance where noninvasive silent speech recognition on an open vocabulary has cleared the threshold of 15\% WER, demonstrating that SSIs can be a viable alternative to automatic speech recognition (ASR). Our work not only narrows the performance gap between silent and vocalized speech but also opens new possibilities in human-computer interaction, demonstrating the potential of cross-modal approaches in noisy and data-limited regimes.
\end{abstract}

\section{Introduction}
\label{introduction}
Silent Speech Interfaces (SSIs) are a branch of human-computer interaction that offers non-invasive means of non-verbal communication. These interfaces may one day be particularly impactful for individuals with speech impairments and in scenarios where traditional vocal communication is impractical or impossible. Despite progress in SSIs, they face significant challenges in achieving sufficiently high accuracy due to both the absence of phonetic content and limited datasets for training and validation.

This research aims to address these limitations by introducing new methodologies that improve silent speech recognition. We propose Multimodal Orofacial Neural Audio (MONA), a novel approach leveraging cross-modal alignment through two new loss functions---cross-contrastive learning (crossCon) and supervised temporal contrastive learning (supTcon). These functions facilitate the training of a multimodal model capable of harnessing audio-only datasets such as LibriSpeech \cite{Panayotov2015}, previously untapped for silent speech recognition.

Additionally, we incorporate Large Language Model (LLM) Integrated Scoring Adjustment (LISA) to significantly improve recognition accuracy. Our methods collectively aim to reduce the word error rate (WER) in silent speech, which is crucial for the practical applicability of SSIs in real-world scenarios.

By narrowing the performance gap between silent and vocalized speech, MONA LISA may help create viable SSI alternatives to existing automatic speech recognition systems. This advancement could enable communication methods for individuals with speech disorders and create a new interface for conversational AI powered by silent speech.

\section{Related work}
\label{related}

SSIs have historically faced challenges such as the absence of phonetic information generated by speech articulators in unrecordable locations and the paucity of training data, impeding their ability to achieve error rates suitable for practical use. Early efforts in SSIs date back to the 1980s, demonstrating successful decoding across a broad array of phonemes with a limited vocabulary using finite automaton \cite{emg1} or maximum likelihood estimation \cite{emg2}. Progress in the field saw a significant leap in the early 2000s with multisession decoding that achieved 87\% accuracy on a vocabulary of 10 words using HMMs \cite{maier-hein}. Jou et al. bootstrapped a silent speech HMM using ASR to 70\% accuracy on a 100-word vocabulary
\cite{Jou2006TowardsCS,Jou2008EARSEA}. \citet{elmahdy} used a deep learning model with convolution, RNN, and CTC loss to achieve 20\% WER on a 20-word vocabulary. Recently, \citet{Gaddy:EECS-2022-68} achieved a breakthrough in open-vocabulary decoding, training a ResNet-Transformer model on an 18-hour dataset to achieve a 29\% WER when predicting text in an open-vocabulary setting, and a 36\% WER when directly synthesizing audio \cite{gaddy-klein-2020-digital,gaddy-klein-2021-improved}. \citet{inproceedings} uses the same dataset and architecture to develop an active-learning paradigm to reduce the labeling burden on EMG data collection.

Technological approaches to SSIs have been diverse, including brain implants \cite{Willett2023,Metzger2023}, lip reading \cite{shi2022learning}, ultrasound \cite{SottoVoce,earssr}, MRI \cite{Tang2023}, fNIRS \cite{Liu2018}, MEG \cite{Dfossez2023}, EEG \cite{eegreview}, radar \cite{Wagner2022}, strain sensors \cite{Kim2022} and non-audible murmur \cite{nakajima2006non}. Among noninvasive techniques, lip reading and surface electromyography (EMG) are notable for their ability to perform high-accuracy open-vocabulary decoding. Lip reading currently shows the best performance in open vocabulary settings when trained extensively, although its accuracy decreases with reduced training data, underperforming EMG trained on fewer hours of data \cite{shi2022learning,Gaddy:EECS-2022-68}. In theory, EMG may have a lower error floor, as non-visible information can be recorded. When placed over facial muscles, EMG can reliably detect activity related to speech articulation \cite{schultz2010}, and when placed on the throat, EMG can detect internal motion of the larynx and vocal cord \cite{Bruder2019}.

Substantially more effort has been devoted to the development of machine learning in automatic speech recognition (ASR) than SSIs, so we look to the machine learning ASR literature for inspiration. Initial progress in ASR came from advances in algorithms, from the introduction of beam search \cite{harpy} and hidden Markov models (HMM) \cite{baker1973machine}, to neural networks \cite{hintonasr} and end-to-end deep learning \cite{deepspeech}. Recently, advances in ASR have predominantly come from new loss functions like InfoNCE \cite{infonce}, unsupervised pretraining as in wav2vec 2.0 \cite{Baevski2020}, and leveraging massively more training data as in Whisper \cite{whisper}. By combining contrastive loss functions for unsupervised training and supervised training on LibriSpeech, w2v-BERT \cite{chung2021w2vbert} achieved a record 2.3\% WER on within-dataset testing.

\section{Problem statement}
\label{problem}

SSIs offer novel communication abilities for people with speech impairments and users in environments where vocal communication is not feasible. SSIs have the potential to restore natural speech in patients with laryngectomy \cite{emg1} or dysarthria and to facilitate seamless and private communication with AI assistants \cite{Kapur2018}. However, these interfaces face inherent challenges due to lack of intelligible sound production. This requires advanced machine learning systems capable of solving a recognition problem that exceeds human capabilities.

The performance threshold for SSIs to become a viable alternative to existing automatic speech recognition (ASR) systems is approximately 15\% WER \cite{Pandey2021}. Despite advances in the field, the challenge remains to improve the accuracy and applicability of SSIs to reach this critical performance threshold. Achieving this level of accuracy is crucial for the advancement of SSI technology and unlocking its potential in a wide range of applications, including silent communication in environments sensitive to noise. Our research focuses on EMG data, given its potential for lower error rates and its ability to record non-visible information related to speech articulation \cite{schultz2010, Bruder2019}.

\section{Approach}
\label{approach}
We introduce two new loss functions for cross-modal contrastive learning, a new latent space alignment approach for silent and vocalized speech using dynamic time warping, and a novel post-processing step following beam search to synthesize the predicted sentence from the top k candidates.

To evaluate our contributions empirically, we utilize the dataset from \citet{gaddy-klein-2020-digital}. The core challenge addressed by Gaddy and Klein was to convert silently mouthed words into audible speech based on EMG sensor data, and so the dataset comprises comprises EMG sensor measurements captured on the face and neck during both vocalized and silently articulated speech. Our methodology closely aligns with the architecture and train / val / test split of \citet{Gaddy:EECS-2022-68}, ensuring consistency in data processing and model evaluation, allowing us to build upon and extend their work in silent speech recognition.

To augment our dataset size, we use LibriSpeech clean + other for training and LibriSpeech clean for validation and test. We remove all chapters from \textit{The War of the Worlds} by H.G. Wells and \textit{The Adventures of Sherlock Holmes} by Arthur Conan Doyle to avoid test label leakage, as these two books are used in our EMG dataset.

Although we focused our efforts on audio and EMG data for this paper, MONA is readily applicable to any pair or group of speech modalities, and LISA is applicable to any speech-to-text prediction model.

\begin{table*}[ht]
\caption{Overview of datasets, data, and corresponding loss functions. $a$ is an audio utterance, $e$ is an emg utterance, and $y$ is the text label for an utterance. For the Gaddy silent dataset, we have parallel readings of a given utterance ($y_1 = y_2$) under both silent and vocalized conditions. A minibatch may contain examples from multiple datasets, and supTcon is applied to all examples.}
\label{table:datasets}
\vskip 0.1in
\begin{center}
\begin{small}
\begin{sc}
\begin{tabular}{lcc}
\toprule
Dataset & Data & Loss Functions \\
\midrule
Gaddy Silent & $(e_1, y_1)$, $(a_2, e_2, y_2)$ & crossCon w/ DTW, supTcon w/ DTW, CTC \\
Gaddy Vocalized & $(a_3, e_3, y_3)$ & crossCon, supTcon, CTC \\
LibriSpeech & $(a_4, y_4)$ & supTcon, CTC \\
\bottomrule
\end{tabular}
\end{sc}
\end{small}
\end{center}
\end{table*}

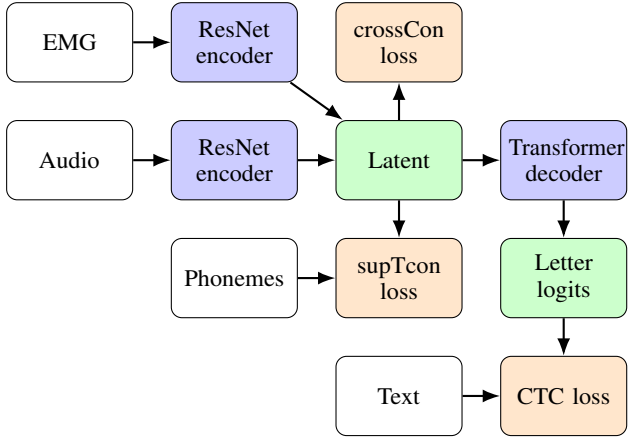
\begin{figure}[ht]
\centering
\begin{tikzpicture}[node distance=0.5cm, auto]
    \node [data] (emg) {EMG};
    \node [block, right=of emg] (resnetEMG) {ResNet encoder};
    \node [data, below=of emg] (audio) {Audio};
    \node [block, right=of audio] (resnetAudio) {ResNet encoder};
    \node [output, right=of resnetAudio] (latent) {Latent};
    \node [block, right=of latent] (transformer) {Transformer decoder};
    \node [output, below=of transformer] (letters) {Letter logits};
    \node [data, below=of resnetAudio] (phonemes) {Phonemes};
    \node [loss, right=of phonemes] (contrastive) {supTcon loss};
    \node [loss, below=of letters] (ctc) {CTC loss};
    \node [data, below=of contrastive] (text) {Text};
    \node [loss, right=of resnetEMG] (crossCon) {crossCon loss};

    \path [line] (emg) -- (resnetEMG);
    \path [line] (resnetEMG) -- (latent);
    \path [line] (audio) -- (resnetAudio);
    \path [line] (resnetAudio) -- (latent);
    \path [line] (latent) -- (transformer);
    \path [line] (transformer) -- (letters);
    \path [line] (letters) -- (ctc);
    \path [line] (latent) -- (contrastive);
    \path [line] (latent) -- (crossCon);
    \path [line] (phonemes) -- (contrastive);
    \path [line] (text) -- (ctc);
\end{tikzpicture}
\caption{MONA architecture.}
\label{fig:mona-architecture}
\end{figure}

\subsection{Cross-contrast (crossCon)}

For vocalized utterance $u$ with simultaneous EMG and audio recordings, let the output of the EMG encoder be
\[\mathbf{Z}_{\text{emg}, u} = \begin{bmatrix} e_{1,u} & e_{2,u} & \cdots & e_{t,u} \end{bmatrix}\]
with $t$ timesteps of 10ms each. The simultaneous output for the audio encoder is
\[\mathbf{Z}_{\text{audio}, u} = \begin{bmatrix} a_{1,u} & a_{2,u} & \cdots & a_{t,u} \end{bmatrix}\]
where $\forall t \forall i (e_{t,u}, a_{t,u})$ are latent embeddings for simultaneously recorded EMG and audio data.

We represent all latent embeddings for the minibatch with $n$ utterances by
\[\mathbf{Z} =  \begin{bmatrix} \mathbf{Z}_{\text{emg}, 1} & \mathbf{Z}_{\text{audio}, 1} & \cdots & \mathbf{Z}_{\text{emg}, n} & \mathbf{Z}_{\text{audio}, n} \end{bmatrix}\]
where $\mathbf{Z} \in \mathbb{R}^{F \times L}$ for $F$ features and $L$ total embeddings. To simplify indexing, we define $i$ as the index for a specific emg frame and the function $j(i)$ to return the corresponding audio index for utterance $u$. Then, crossCon is defined by Equation \ref{eq:crossCon} for temperature $\tau$:

\begin{align}
\mathrm{sim}(\mathbf{z}_i, \mathbf{z}_j; \tau=0.1) &= \exp\left(\frac{\cos(\mathbf{z}_i, \mathbf{z}_j)}{\tau}\right) \label{eq:sim}\\
\mathcal{L}^\mathrm{cross}_i & = -\log\left( \frac{\text{sim}(\mathbf{z}_i, \mathbf{z}_{j(i)})}{\sum_{j \neq i}^{L} \text{sim}(\mathbf{z}_i, \mathbf{z}_j)} \right) \label{eq:crossClass}\\
\mathcal{L}^\mathrm{cross} & = \frac{1}{L} \sum_{i=1}^{L} \mathcal{L}^\mathrm{cross}_i \label{eq:crossCon}
\end{align}

\begin{figure}
    \centering
    \includegraphics[width=0.65\linewidth]{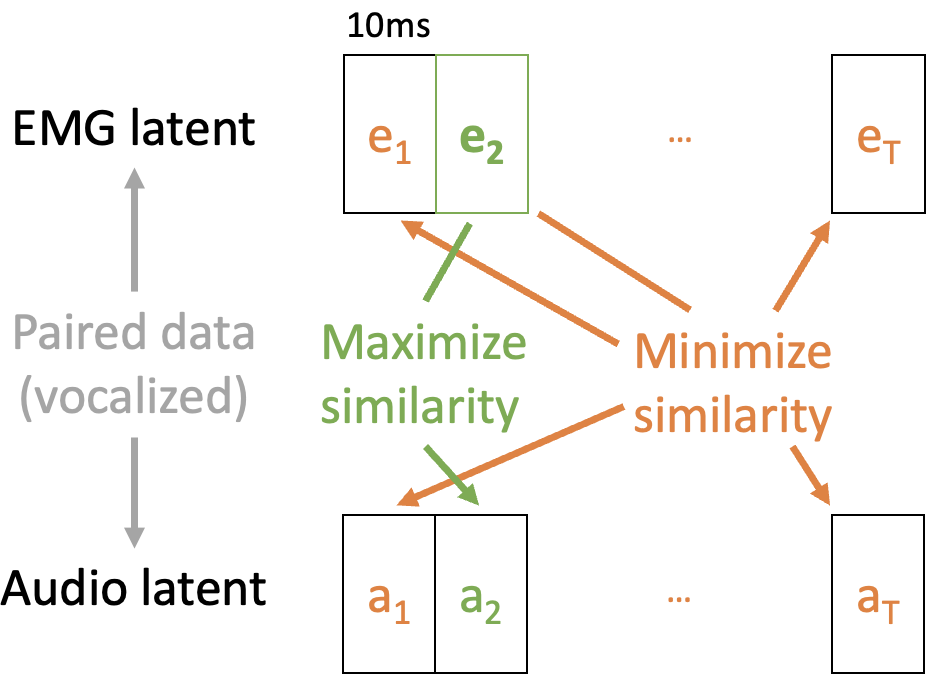}
    \caption{\textbf{Latent space alignment for time-aligned data with crossCon}. The similarity of latent embeddings for simultaneously recorded EMG and audio data is maximized for the same timestep, and minimized to other timesteps.}
    \label{fig:crossCon}
\end{figure}

Equation \ref{eq:crossClass} can be thought of as a negative log-likelihood loss, where we attempt to classify the specific positive pair of EMG \& audio latents among a set of distractors. Equation 3 can be thought of as the average of loss from $L$ classification problems. This differs from CEBRA \cite{Schneider2023} in that we source distractors from both domains of data. Our formulation encourages the difficult task of creating encoders where cross-modality $e_{t,u}$ and $a_{t,u}$ are more similar than the same-modality $e_{t-1,u}$ and $e_{t,u}$, and therefore encourages a shared latent representation for EMG and audio data (Figure \ref{fig:crossCon}). 

\begin{figure}
    \centering
    \includegraphics[width=0.65\linewidth]{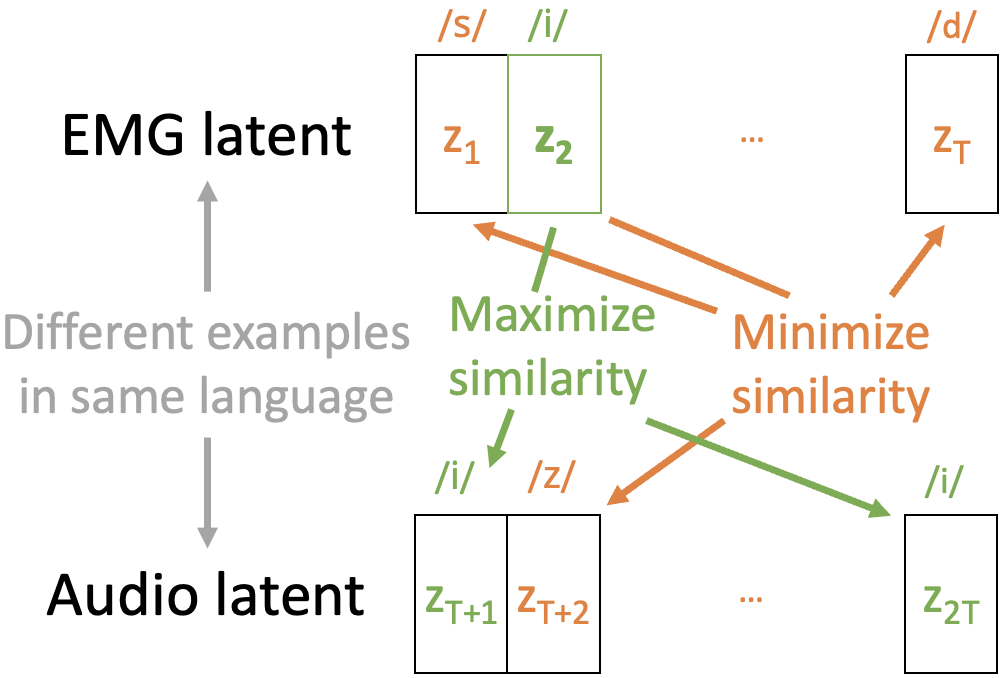}
    \caption{\textbf{Latent space alignment for independent data with supTcon}. The similarity of latent embeddings with the same phoneme class \textbf{/i/} is maximized, and minimized for latent embeddings with different phonemes.}
    \label{fig:supTcon}
\end{figure}

\subsection{Supervised temporal contrast (supTcon)}

supTcon is suitable when data from different modalities or datasets are not acquired synchronously. We use a class label per latent embedding to align latent representations across data modalities and datasets (Figure \ref{fig:supTcon}). Here, the class label is a phoneme or silence as found with Montreal Forced Aligner. To simplify the indexing notation, we introduce the function $p(i)$, which is defined to return the set of all indices corresponding to entries that share the same class label (phoneme) as the entry at index $i$.

\begin{align}
\mathcal{L}^\mathrm{sup}_i & = - \sum_{q \in p(i)} \frac{1}{|p(i)|} \log\left( \frac{\text{sim}(\mathbf{z}_i, \mathbf{z}_q)}{\sum_{j \neq i}^{L} \text{sim}(\mathbf{z}_i, \mathbf{z}_j)} \right) \label{eq:supConSum} \\
\mathcal{L}^\mathrm{sup} & = \frac{1}{L} \sum_{i=1}^{L} \mathcal{L}^\mathrm{sup}_i
\end{align}

Following the analysis in \citet{supconlearn}, we perform the summation over positives in Equation \ref{eq:supConSum} outside of the log. However, in supTcon, the number of comparisons is quadratic in the total number of timesteps across all examples in the batch, rather than quadratic in the number of examples per batch.

\subsection{supTcon and crossCon for silent speech with Dynamic Time Warping (DTW)}
Since the silent speech data does not contain usable audio, we take advantage of the parallel silent and vocalized utterances in \citet{gaddy-klein-2020-digital} to align activity across conditions. Since DTW on EMG data performs worse than on audio data \cite{Gaddy:EECS-2022-68}, we instead temporally align the silent condition with the vocalized condition in the latent space. This allows the encoder to act as a denoiser and extract meaningful semantic content that may improve DTW results. Together, we call this approach MONA (Figure \ref{fig:mona-architecture}).

Let the emg latent of silent utterance be $Z_1 \in \mathbb{R}^{F \times T_1}$ with $F$ features and $T_1$ timesteps, and let parallel vocalized emg recording of utterance be $Z_2 \in \mathbb{R}^{F \times T_2}$ and phonemes $P_2 \in \mathbb{P}^{T_2} $ where $\mathbb{P}$ is set that includes phonemes and silence token with $T_2$ timesteps. We first calculate the Euclidean distance matrix $D \in \mathbb{R}^{T_1 \times T_2}$. Then, using DTW, we calculate the alignment of $Z_2$ onto $Z_1$, and use this alignment to create the warped $\hat{Z}_2 \in \mathbb{R}^{F \times T_1}$ and $\hat{P}_2 \in \mathbb{P}^{T_1}$. We can now calculate crossCon and supTcon for the Gaddy silent dataset.

\subsection{Greedy Class-Weighted Bin Packing}
In order to calculate crossCon or supTcon + DTW, we must have at least one silent EMG example with parallel vocalized EMG and Audio per minibatch. As memory usage for contrastive loss functions are quadratic with input length, we must additionally sample batches below a maximum length to ensure sufficient GPU memory. Finally, since we have massively more audio-to-text data than emg-to-text, we choose to undersample LibriSpeech so that it accounts for approximately 50\% of the examples per epoch. Our heuristic for solving this class-proportional bin packing problem is described in Algorithm \ref{alg:binpacking}, and is used for our Batch Sampler.

\subsection{LLM-integrated scoring adjustment (LISA)}\label{lisa-technique}

During training, we use a 4-gram language model with a beam size of 150 for beam search. By increasing the beam size to 5000 for inference, we reduce WER by ~1\%. Manual inspection of the top 100 beams revealed that after viewing the list of predictions, in some cases a human could succeed in writing down the correct sentence despite this text not appearing exactly as any specific prediction. As such, rather than calculating the posterior negative log-likelihood (NLL) from the beam search prior and updating with a LLM \cite{Willett2023}, we instead prompt GPT-3.5 (\citet{gpt3}, ``gpt-3.5-turbo-16k-0613") or GPT-4 (``gpt-4-0125-preview") with a task description and a list of the top predictions. These predictions can be sourced either from the top $k$ beams from beam search or from an ensemble of models. The latter has a higher diversity of predicted text and is depicted below.

\begin{quotation}
Your task is to perform automatic speech recognition. Below are multiple candidate transcriptions, listed from most likely to least likely. Choose the transcription that is most accurate, ensuring it is contextually and grammatically correct. Focus on key differences in the options that change the meaning or correctness. Avoid selections with repetitive or nonsensical phrases. In cases of ambiguity, select the option that is most coherent and contextually sound. Respond with the chosen transcription only, without any introductory text. \newline
after breakfast instead forking at aside to walk down towards the common\newline
after breakfast stead of working a decided to walk down towards the common\newline
after breakfast stead working a decided to walk down towards the common\newline
after breakfast instead working a sudden to walk down towards the common\newline
after breakfast instead of working i decided to walk down towards the common\newline
after breakfast instead of working a decided to walk down towards the common\newline
after breakfast instead of working a descended to walk down towards the common\newline
after breakfast instead of working at a sudden to walk down towards the common\newline
after breakfast instead of working a decided walk down towards the common\newline
after breakfast instead of working a decided to walk down towards the common
\end{quotation}

This prompt (``Ensemble 10 x top 1") results in the text shown in Table \ref{tab:transcription_examples}, where LISA corrects all four errors present in this utterance.

\subsection{Experiment setup}
We dynamically construct minibatches so that at least one Gaddy silent example (Table \ref{table:datasets}) is present in each minibatch, and each minibatch is class-balanced for Gaddy silent and Gaddy vocalized, with examples from LibriSpeech filling the remaining 50\% of examples. This ensures that each gradient update integrates the loss of each dataset, jointly optimizing for both domains. crossCon and supTcon are quadratic in memory usage in relation to total number of timesteps across the batch, with the latter permitting up to 128K timesteps into the 80GB of VRAM for a Nvidia A100 80GB. With crossCon, up to 256K timesteps were possible on a single GPU.

In an effort to improve numerical stability, we use GeLU activations for the ResNet encoder without pre-norm activation, and, taking inspiration from \citet{balduzzi17b}, we scale the residual path of each block in the ResNet encoder by $\frac{1}{\sqrt{2}}^\ell$ where $\ell$ is the layer number. To increase training speed, we leverage the TensorFloat32 format.

We use the same random number generator seed for all data loaders to ensure that each model sees the same sequence of batches during training. An epoch ends when there are no more available samples for a given class, meaning that each epoch typically includes all examples from Gaddy but only a fraction from LibriSpeech. Each model is trained five times with different random initializations. We select the model with the best validation WER on silent EMG for MONA, and ensemble the best 10 models as ranked by validation WER on silent EMG for MONA LISA. For fine-tuning LISA, we split validation into two halves of 100 examples each, fine-tuning on the first 100 and evaluating on the second 100.

A first draft of this manuscript was produced, including all figures, before the evaluation of the test data. We selected the best MONA and MONA LISA models on the basis of the validation WER for silent EMG. The values reported in Table \ref{tab:wer_comparison} reflect the test performance of our models, as chosen \textit{a priori} to the first evaluation on the test set.

\section{Results}

\begin{figure*}
    \centering
    \includegraphics[width=1.0\textwidth]{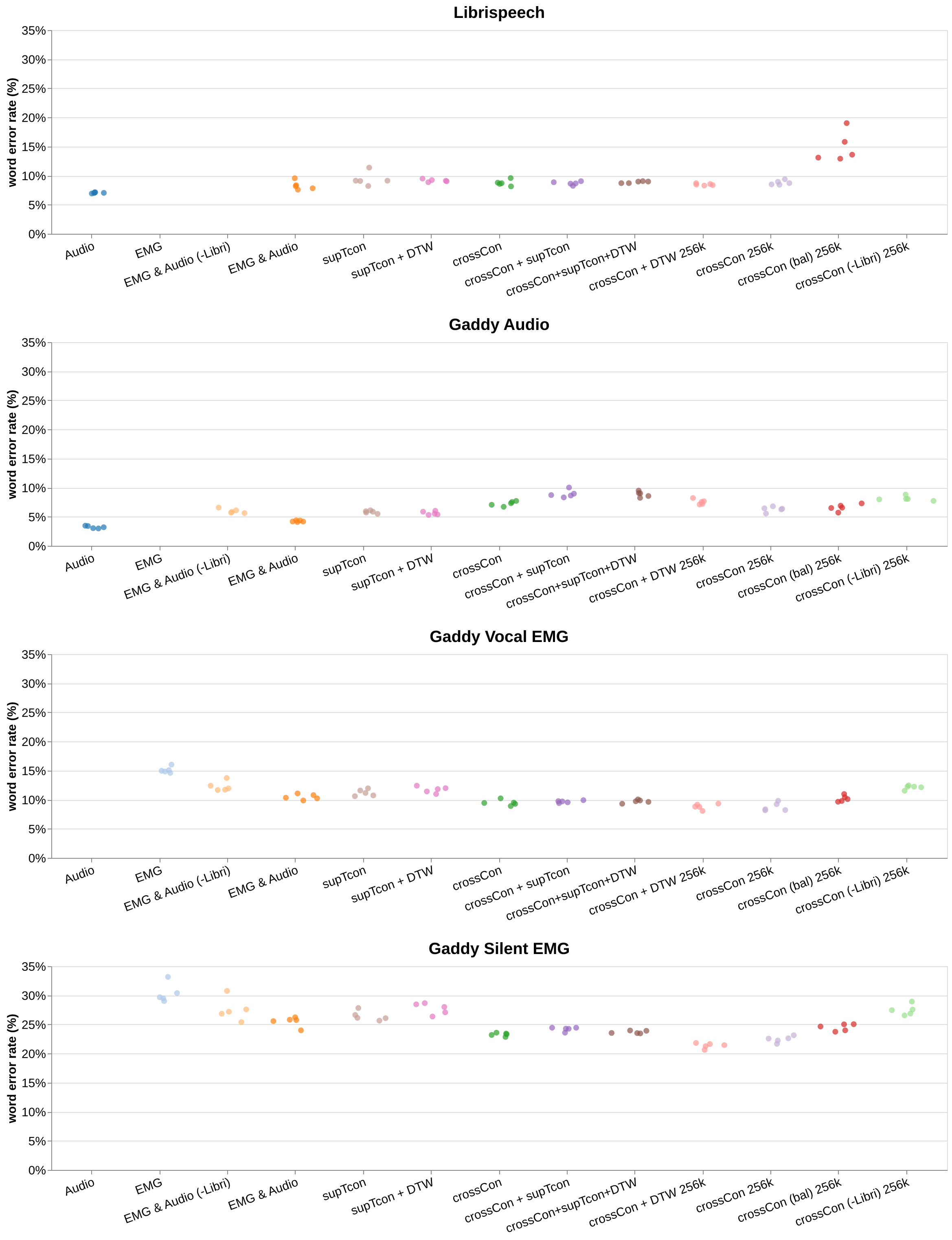}
    \vskip -0.1in
    \caption{Comparison of validation WER for the MONA architecture with varying loss function. -Libri = No LibriSpeech, bal = Balanced audio/EMG sampling.}
    \label{fig:mona-wer}
\end{figure*}

To decrease measurement error in comparing model performance, in this section we report the average WER on silent EMG validation from 5 models trained with different initial seeds unless otherwise noted. As shown in Figure \ref{fig:mona-wer}, our baseline model performs moderately worse on silent speech validation data (30.4\% average WER) than the model by \citet{Gaddy:EECS-2022-68} (28.8\% WER). Beyond our changes to improve numerical stability and training speed, the addition of audio data may additionally negatively impact batch norm statistics in the decoder despite $\lambda_{audio} = 0$ in the loss function (see Appendix \ref{app:mona-loss}), leading to degraded decoder performance. However, our baseline model performs significantly better on the vocal EMG validation data, achieving a 15.1\% WER. This improvement may result from the decision to include silent EMG in the training corpus, whereas \citet{Gaddy:EECS-2022-68} trains on vocal EMG only to reach 23.3\% WER.

We observed higher performance in all model formulations when training on batches with up to 256,000 time steps. However, our implementation of supTcon could only fit on a single A100 with up to 128,000 time steps, so we evaluated most loss functions with this shorter batch length for a fair comparison (Figure \ref{fig:mona-wer}). 

The addition of CTC loss on the Gaddy audio data reduces the silent EMG WER to 27.6\%, and incorporating the LibriSpeech Clean \& Other datasets result in a significant reduction in WER to 25.5\%, surpassing both our baseline model as well as the current state-of-the-art result of \citet{Gaddy:EECS-2022-68}. The benefits of the EMG \& Audio model were more pronounced in the vocal EMG validation set, with the WER dropping to 10.5\%.

Latent representations from the Gaddy and LibriSpeech datasets are effectively aligned by crossCon, enhancing the model's robustness in diverse speech scenarios. The integration of crossCon further refined our model's performance, reaching 23.3\% WER on silent EMG.  Without inclusion of the LibriSpeech dataset, crossCon provides no benefit over the EMG + Audio model (Figure \ref{fig:mona-wer}). Compared to the baseline model, a model with crossCon trains substantially faster. This also corresponds to improved generalization: at epoch 30, the baseline model has a 58.3\% WER on validation data vs 43.4\% WER for the model with crossCon (Figure \ref{fig:wer_ctc}). With 256k timesteps per batch, the performance improved further to 22.4\%

As crossCon is only applied to the Gaddy dataset, we sought to formulate a contrastive loss function that could directly align EMG and LibriSpeech data. We anticipated that supTcon might therefore improve performance, but found a modest performance degradation in all models trained for this paper (Figure \ref{fig:mona-wer}).

In order to align silent EMG and audio data, we implemented DTW in latent space such that we can warp parallel vocalized audio latents to silent EMG latents, and apply crossCon and/or supTcon. The combination of crossCon + DTW achieved a 21.4\% average WER on silent EMG. We selected the best model, with a silent EMG validation WER of 20.6\% as MONA.

The largest improvement came from the addition of more powerful language models. The integration of an LLM to rescore the top 10 beams of the best MONA model (``Direct top 10 beams"; Table \ref{tab:rescorings_comparison}) corrected for a significant fraction of remaining errors, bringing the validation WER to 18.0\%. We noticed that the variability of predictions from beam search was relatively constrained, with most beams differing by only one word, whereas the predictions from models with different initialization varied substantially. By passing the predictions from an ensemble of 10 different MONA models through an LLM, we achieve a 13.1\% WER. Finally, by fine-tuning the LLM, we reach the lowest validation WER of 7.3\% (Table \ref{tab:rescorings_comparison}). We call this method of passing an ensemble of 10 MONA models through a fine-tuned LLM, ``MONA LISA".

MONA LISA generalizes well to the test set, achieving a record 12.2\% WER on silent EMG, and 3.7\% WER on vocal EMG (Table \ref{tab:wer_comparison}), reducing the state-of-the-art WER by 57.6\% and 84.1\% respectively.

Finally, we applied LISA to the \href{https://eval.ai/web/challenges/challenge-page/2099/overview}{Brain-to-Text Benchmark `24}. The dataset consists of 12,100 sentences of intended speech by an individual with late-stage Amyotrophic Lateral Sclerosis (ALS), while neural activity was recorded from 256 electrodes in speech-related areas of motor cortex. We fine-tuned LISA on an ensemble of 10 $\times$ LSTM, running beam search using a 5-gram language model, using the \href{https://github.com/cffan/neural_seq_decoder}{Pytorch implementation} of the model from \citet{Willett2023}. At time of writing, LISA is the top-ranked model on the leaderboard, reducing the top WER from 9.8\% to 8.9\%.

\begin{table}[htbp]
\caption{Example validation transcription before and after LISA}
\label{tab:transcription_examples}
\vskip 0.1in
\begin{center}
\begin{small}
\begin{sc}
\begin{tabular}{lp{0.7\linewidth}}
\toprule
Method & Transcription \\
\midrule
Beam Search & after breakfast instead [of] \textbf{forking at aside} to walk down towards the common \\
LISA & after breakfast instead of working i decided to walk down towards the common \\
Actual & after breakfast instead of working i decided to walk down towards the common \\
\bottomrule
\end{tabular}
\end{sc}
\end{small}
\end{center}
\end{table}

\begin{table}[t]\label{test-wer}
\caption{Test-set word error rate (WER) comparison of different models and enhancements on the Gaddy 2020 benchmark.}
\vskip 0.1in
\centering
\begin{small}
\begin{sc}
\begin{tabular}{lcccc}
\toprule
 & Libri- & Gaddy & Vocal & Silent \\
Model & Speech & Audio & EMG  &  EMG \\
\midrule
\citet{Gaddy:EECS-2022-68} & - & - & - & 28.8\% \\
\citet{Gaddy:EECS-2022-68} & - & - & 23.3\% & - \\
\citet{Gaddy:EECS-2022-68} & - & 11.3\% & - & - \\
Whisper v2 & \textbf{2.7\%} & \textbf{2.3\%} & - & - \\
MONA & 9.1\% & 7.7\% & 8.9\% & 22.2\% \\
MONA LISA & 5.7\% & 2.6\% & \textbf{3.7\%} & \textbf{12.2\%} \\
\bottomrule
\end{tabular}
\end{sc}
\end{small}
\label{tab:wer_comparison}
\end{table}

\begin{table}[ht]
\caption{Validation WER of LISA approaches on silent EMG data.}
\vskip 0.1in
\centering
\begin{small}
\begin{sc}
\begin{tabular}{lcc}
\toprule
Approach & GPT-3.5 & GPT-4 \\
\midrule
Chain of Reasoning & 19.5\% & 19.2\% \\
Direct top 100 beams & 18.0\% & 16.9\% \\
Direct top 10 beams & 18.0\% & 17.6\% \\
Ensemble 10 $\times$ top 10 & 13.0\% & 14.4\% \\
Ensemble 10 $\times$ top 1 & 13.1\% & 13.9\% \\
fine-tuned 15 $\times$ top 1 & 7.6\% & - \\
fine-tuned 10 $\times$ top 1 & 7.3\% & - \\
\bottomrule
\end{tabular}
\end{sc}
\end{small}
\label{tab:rescorings_comparison}
\end{table}

\section{Discussion}
\label{discussion}
Our approach contrasts with existing methods in SSIs by heavily relying on audio data from vocal utterances to enhance the decoding of silent speech. We also utilize large language models (LLMs) to capitalize on extensive text datasets to improve silent speech recognition. These strategies have significantly narrowed the performance gap between audio and silent sEMG from 26.5\% to 9.9\% in absolute percentage points. Notably, our work has achieved a WER below the critical 15\% threshold, indicating the viability of SSIs for open-vocabulary applications.

Compared to the baseline model, a model with crossCon trains substantially faster. This also corresponds to an improved generalization: At epoch 30, the baseline model has a 58.3\% WER on validation data vs 43.4\% WER for the model with crossCon (Figure \ref{fig:wer_ctc}). When developing crossCon, we also explored other formulations for encouraging alignment. A simple MSE on paired latent embeddings moved latents towards zero while harming performance. If dot product is used instead of cosine for Equation \ref{eq:sim}, then the magnitude differs for the latents of each domain, and there are no performance gains.

We observed that CTC loss is only loosely predictive of WER; validation CTC loss typically reached its minima around epoch 40-60, while validation WER reached it's minima around epoch 125-200 (Figure \ref{fig:wer_ctc}). This phenomena has been previously observed for Deepspeech2 \cite{SiddGururani2017} and Conformers as implemented in Nvidia NEMO \cite{psydok2022}, indicating that these model may become more confident in its predictions (and thus have a lower WER), even as the overall probability distribution modeled by the network, reflected in the CTC loss, becomes less aligned with the true distribution of the validation data. This may indicate that overfitting helps the approximate beam search algorithm find a likely answer with higher reliability. Future research might explore if LISA works better on checkpoints with low CTC loss, or low WER. The overfitting of CTC also invites research that explores new sequence-to-sequence loss functions that generalize without mode collapse.

Although our core contributions held up in the final evaluation on test data, the benefit from fine-tuning on GPT-3.5-turbo generalized poorly on silent EMG data. We tried fine-tuning GPTs on multiple different ensembles: (a) the 10 best models of Figure \ref{fig:mona-wer}, 5 $\times$ (crossCon + DTW 256k) and 5 $\times$ (crossCon 256k), (b) we trained an additional 5 $\times$ (crossCon + DTW 256k) and selected the 9 $\times$ (crossCon + DTW 256k) and 1 $\times$ (crossCon 256k) with the lowest validation WER, and (c) the top 15 models 10 $\times$ (crossCon + DTW 256k) and (5 $\times$ crossCon 256k). Since the fine-tuned model on (a) had the lowest WER on the witheld validation utterances, we selected that model for final test set evaluation per our selection criteria.

In Appendix \ref{app:lisa-test}, we examine the performance of these ensembles as well as other fine-tuning methods, and speculate that alternate construction and selection criteria may be warranted when fine-tuning an LLM for LISA. Fine-tuned ensemble (b) achieved an 8.0\% WER on silent EMG test data, and fine-tuned ensemble (c) achieved a 9.1\% WER, while fine-tuned ensemble (a) recorded a 12.2\% WER. Fine-tuning an ensemble on audio predictions from Librispeech also resulted in $<$10\% WER. We hypothesize that fine-tuning on a diverse dataset may encourage task performance through ensemble weighting, whereas fine-tuning within-domain on limited examples carries the risk of overfitting to the lexicon of the provided predictions.

These learnings informed our training approach for LISA on the Brain-to-text `24 competition. We fine-tuned the 10 $\times$ top 1 model on all 600 examples of the ``test" data (we consider this the validation set) before evaluating on the 1200 utterances in the ``competition" data (we consider this the test set). LISA achieves 13.8\% WER on validation prior to fine-tuning, and 10.4\% WER on validation after fine-tuning (train / evaluation on same samples). The fine-tuned model generalizes well, reaching a record 8.9\% WER on the held-out competition data.

One challenge for deploying LISA in real-time inference settings is the 10x increase in compute required to obtain ensemble predictions. Future work might explore ensemble approximation methods like sampling multiple predictions from a single model with dropout \cite{gal2016dropout}, or acquiring multiple predictions using a mixture of experts \cite{MOE}. One additional challenge is the instability of ChatGPT API results overtime, which may require new prompt engineering to maintain performance. Future work might explore the use of an open source model such as LLaMA 2 \cite{2023llama} or Mixtral 8x7B \cite{jiang2024mixtral} for stable performance with long-term reproducibility.

During development, supTcon provided modest improvement under a different set of hyperparameters than ulitimately used in this paper. The approach may warrant continued exploration due to its potential to be used when only silent EMG data are available. Here, we only consider supervised learning where text labels are available; however, supTcon might be extended to the semi-supervised learning case where only subsets of class labels are available, perhaps by using gumbel softmax for phoneme classification instead of DTW. This allows for training on additional data, which may be particularly useful when SSIs are worn outside of the lab.

Inclusion of a text modality during training may further boost performance, but requires a phoneme duration prediction or other alignment approach, similar to the duration / pitch predictor in NaturalSpeech 2 \cite{shen2023naturalspeech}. Additional context in the form of longer utterances (e.g. 30s) or text transcripts of previous utterances may further allow for LISA to improve imputation of missing phonetic content.

\section{Conclusion}
\label{conclusion}
The present study demonstrates the effectiveness of cross-modal training through novel loss functions and a new latent space alignment approach. By further leveraging LLMs for scoring adjustment, we have significantly reduced the WER in silent speech recognition. Future work might explore the application of these techniques to additional speech modalities, such as invasive BCI or next-generation SSIs.

Our work significantly narrows the performance gap between silent and vocalized speech. This not only illustrates the feasibility of high-accuracy SSIs but also opens new avenues in human-computer interaction, particularly for individuals with speech impairments and in scenarios where traditional vocal communication is impractical. The findings suggest that the performance gap between silent EMG data and ASR for open vocabulary may yet be closed for a single speaker with sufficient electrode coverage or a combination of multiple SSI modalities.

\begin{algorithm}[H]
   \caption{Greedy Class-Weighted Bin Packing}
   \label{alg:binpacking}
\begin{algorithmic}
   \STATE {\bfseries Input:} item lengths \( L \), item class labels \( C \), max bin length \( M \), class proportions \( P \), required classes per bin \( I \)
   \STATE {\bfseries Output:} list of bins with item indices \( B \)
   \STATE
   \STATE {\bf Initialize} \( B \) and sum list \( sums \) to empty
   \STATE {\bf Group} indices by class into \( idx \)
   \STATE {\bf Shuffle} \( idx \) per class
   \STATE {\bf Initialize} \( debt \) for each class to 0
   \WHILE{num(\( idx \)) \( > \) 0 for all classes}
       \STATE {\bf Sample} class \( c \) based on \( P \)
       \IF{\( debt \) of \( c \) \( > \) 0}
           \STATE {\bf Decrement} \( debt \) for \( c \)
           \STATE {\bf Continue}
       \ENDIF
       \STATE {\bf Pop} item with length \( \ell \) from \( L \) and class \( c \) from \( C \)
       \IF{there exists a bin where \( \ell \) + sum(lengths in bin) \( \leq M \)}
           \STATE {\bf Find} the first such bin
           \STATE {\bf Add} item \( idx \) to this bin and update \( sums \)
       \ELSE
           \STATE {\bf Create} new bin; {\bf Add} item \( idx \) to bin
           \IF{\( I \) classes are missing}
               \STATE {\bf Add} \( idx \) for missing \( I \) classes to new bin
               \STATE {\bf Increment} \( debt \) for added classes
           \ENDIF
       \ENDIF
   \ENDWHILE
   \STATE {\bf Initialize} failure count \( f \) to 0
   \STATE {\bf Set} failure threshold \( T \)
   \WHILE{failure count \( < T \) and remaining items exist}
       \STATE {\bf Adjust} \( P \) to avoid sampling classes without items
       \STATE {\bf Sample} class \( c \) based on \( P \)
       \STATE {\bf Try} to add item to existing bin without exceeding \( M \)
       \IF{added successfully}
           \STATE {\bf Reset} \( f \)
       \ELSE
           \STATE {\bf Increment} \( f \)
       \ENDIF
   \ENDWHILE
   \STATE {\bf Discard} any remaining items
\end{algorithmic}
\end{algorithm}
\vspace{-0.5cm}
\section*{Software and Data}

All software is available at \url{https://github.com/tbenst/silent_speech}. Data is available at \url{https://zenodo.org/records/4064409}. Training logs and metrics are available for all runs at: \urlhref{https://app.neptune.ai/o/neuro/org/Gaddy/runs/table?viewId=98bd0377-3fc9-4df2-a53c-1ee50dddb2c5}{https://app.neptune.ai/o/neuro/org/Gaddy}

\section*{Impact Statement}
This paper presents work with the aim of advancing machine learning techniques for SSIs. Potential societal consequences of our work are numerous, including voice restoration for patient populations with speech disorders, invisible computer interaction, and the decoding of subvocalizations. Decoding inner speech does not appear to be possible with EMG \cite{Nalborczyk2020}, so the prospects of using this work to coercively record private thoughts appear remote. Therefore, similar ethical considerations apply as with other speech decoding technology, like ASR.

\section*{Conflict of Interest}
In the interest of full transparency, we disclose the following: A preliminary patent application (USPTO No. 63/588,649) related to the work presented in this paper has been filed by
Tyler Benster, Shaul Druckmann, Reshef Elisha, and Guy Wilson by Stanford University.
Additionally,
Tyler Benster, Reshef Elisha, and Guy Wilson
have founded a startup company aiming to commercialize technologies related to the research discussed herein.

\section*{Acknowledgements}

This work was supported by a seed grant from
the Stanford Institute for Human-Centered Artificial Intelligence (HAI) and the Wu Tsai Neurosciences Institute. Additional funding was provided by the National Institutes of Health under Grant numbers R01-EB028171 and U01-DC019430, along with support from the Stanford Center for Mind, Brain, Computation, and Technology.
We extend our gratitude for the funding and resources that made this research possible. Special thanks to 
Jaimie Henderson, Steve Baccus, Scott Linderman, Yeongjun Lee, Zhenan Bao, Jeray Thelwell, Winson Cheng, and Arlo Faria
for their valuable comments and insights.

\bibliography{references}
\bibliographystyle{icml2024}

\newpage
\appendix
\onecolumn
\section{Appendix}

\subsection{MONA loss function}\label{app:mona-loss}
Let $\mathcal{L}^\text{emg}$ and $\mathcal{L}^\text{audio}$ be the Connectionist Temporal Classification (CTC) loss \cite{Graves2006} for EMG and audio data, respectively. 
We define the loss for MONA as follows: 

\begin{align*}
\mathcal{L} = \ & \lambda_{\text{emg}} \cdot \mathcal{L}^{\text{emg}} \\
             & + \lambda_{\text{audio}} \cdot \mathcal{L}^{\text{audio}} \\
             & + \lambda_{\text{cross}} \cdot \mathcal{L}^{\text{cross}} \\
             & + \lambda_{\text{sup}} \cdot \mathcal{L}^{\text{sup}}
\end{align*}

For the experiments in Figure \ref{fig:mona-wer}, we use the following values of $\lambda$:

\begin{table}[h]
\caption{Configuration of lambda parameters for different model runs.}
\vskip 0.1in
\centering
\begin{tabular}{lcccc}
\hline
\textbf{Model} & $\lambda_{\text{audio}}$ & $\lambda_{\text{emg}}$ & $\lambda_{\text{sup}}$ & $\lambda_{\text{cross}}$ \\
\hline
Audio              & 1   & 0    & 0     & 0  \\
EMG                & 0   & 1    & 0     & 0  \\
Audio \& EMG       & 1   & 1    & 0     & 0  \\
supTcon            & 1   & 1    & 0.1   & 0  \\
crossCon           & 1   & 1    & 0     & 1  \\
crossCon + supTcon & 1   & 1    & 0.1   & 1  \\
\hline
\end{tabular}
\label{tab:lambda_config}
\end{table}

\begin{figure}
    \centering
    \includegraphics[width=0.9\linewidth]{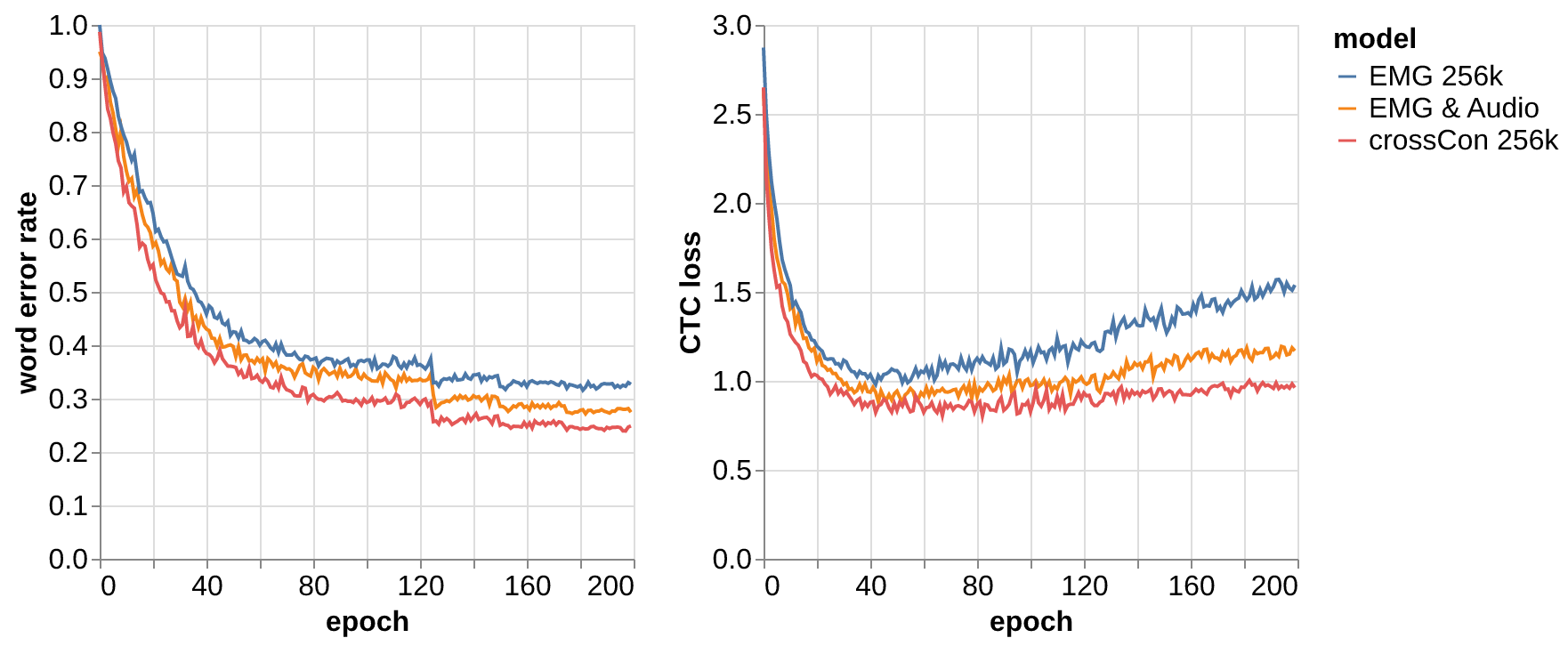}
    \caption{Median of five runs WER and CTC loss by training epoch on validation data using 150 beams.}
    \label{fig:wer_ctc}
\end{figure}

\subsection{Balanced EMG/Audio sampling}
Both the EMG-only model and the supTcon model suffered from NaN issues during training. We hypothesized that this came from the presence of minibatches with few EMG utterances. To address, we calculated class sampling proportions that in expectation lead to the same number of EMG \& Audio utterances per batch with the silent:vocalized EMG dataset ratio: 18.3\% silent + parallel EMG, 18.3\% LibriSpeech, and 63.3\% vocalized EMG (compared with 11.2\%, 50\%, and 38.8\% for all other experiments). This indeed stabilized the gradients for the supTcon model, allowing it to train with (max\_len=128000, grad\_accum=2) instead of (max\_len=128000, grad\_accum=4) or (max\_len=256000, grad\_accum=2), but led to similar validation WER. However, the EMG-only model failed to train with (max\_len=128000, grad\_accum=2), still requiring the latter two configurations and similarly seeing no change in validation WER. For crossCon, balanced sampling hurt performance on all tasks. We hypothesize that the training stability benefits of balanced EMG/audio sampling are outweighed by the decrease in performance by undersampling LibriSpeech.

\subsection{Test performance of fine-tuned LISA}\label{app:lisa-test}

\begin{table}[ht]
\caption{Silent EMG validation and test WER of LISA for the three evaluated ensembles. Ensembles were created and fine-tuned prior to test, and model in bold was chosen based on validation performance.}
\vskip 0.1in
\centering
\begin{small}
\begin{sc}
\begin{tabular}{lcccc}
\toprule
 & \multicolumn{2}{c}{No fine-tuning} & \multicolumn{2}{c}{fine-tuning} \\
\cmidrule(r){2-3} \cmidrule(r){4-5}
Ensemble for LISA & Validation & Test & Validation & Test \\
\midrule
\textbf{5 $\times$ (crossCon + DTW), 5 $\times$ crossCon} & 13.3\% & 13.2\% & \textbf{7.3\%} & 12.2\% \\
10 $\times$ (crossCon + DTW), 5 $\times$ crossCon & 12.0\% & 12.7\% & 7.6\% & 9.1\% \\
9 $\times$ (crossCon + DTW), 1 $\times$ crossCon & 11.9\% & 12.8\% & 7.6\% & 8.0\% \\
\bottomrule
\end{tabular}
\end{sc}
\end{small}
\label{tab:lisa_test}
\end{table}

If we fine-tune the first ensemble in Table \ref{tab:lisa_test} with ensemble audio predictions from half of the LibriSpeech validation set (270 examples), we achieve a Test performance of 9.2\% on silent EMG. Based on these results, we hypothesize that fine-tuning to the silent EMG validation set, where all 100 examples come from \textit{War of the Worlds} may have a higher risk of overfitting to the particular vocabulary in those examples rather than fine-tuning to the larger and more diverse vocabulary in LibriSpeech, which may encourage the LLM to focus on the task of weighting the different models.

Although the fine-tuning results generalize well from dataset-to-dataset for a particular ensemble, they do not generalize well from ensemble-to-ensemble. The test WER is 15.0\% using the 10 $\times$ ensemble Librispeech fine-tune on the 15 $\times$ ensemble predictions (using the same 10 with an additional 5 models), and 21.1\% when using a different set of 10 models (row 3 of Table \ref{tab:lisa_test}, despite the new set consisting of models with lower average validation WER (21.4\% vs 21.9\%). This supports our hypothesis that fine-tuning on Librispeech is largely learning to weight the predictions of the different models in the ensemble, rather than learning the word statistics of the particular dataset.

\subsection{Alternate LISA prompts}
\subsubsection{Chain of Reasoning}
\begin{quotation}
Your task is to perform automatic speech recognition. Below are multiple candidate transcriptions, listed from most likely to least likely. Begin your response with a Chain of Reasoning, explaining your analysis and decision-making process in choosing the most accurate transcription. After your analysis, clearly indicate your final choice with the cue `TRANSCRIPT: '. Ensure the transcription you choose is contextually and grammatically correct. Focus on key differences in the options that change the meaning or correctness. Avoid selections with repetitive or nonsensical phrases. In cases of ambiguity, select the option that is most coherent and contextually sound. Respond first with your reasoning, followed by `TRANSCRIPT: ' and then the chosen transcription."
\end{quotation}
The model exhibited ~3\% noncompliance to the task on both GPT-3.5 and GPT-4, either refusing to make a selection or failing to respond with ``TRANSCRIPT:". We excluded these predictions from Table \ref{tab:rescorings_comparison}. Without excluding, the WER increases to $>$30\%. The poor performance of chain of reasoning is surprising given the success of this technique in a wide variety of other tasks. We hypothesize that the chain of reasoning may inject the model's own lexicon statistics into the task, and thereby detract from the lexicon statistics of the predictions at hand.
\subsubsection{NLL Loss}
\begin{quotation}
Your task is automatic speech recognition.
Below are the candidate transcriptions along with their
negative log-likelihood from a CTC beam search.
Respond with the correct transcription,
without any introductory text.
\end{quotation}
We found that including NLL losses leads to worse performance than a direct approach excluding the phrase ``along with their negative log-likelihood from a CTC beam search" and the corresponding NLL values. We hypothesize that the LLM is not capable of doing a Bayesian update by multiplying the token-encoded NLL with its own internal logits and so these numbers are more distracting than simply ranking from best to worst.

\subsection{Test performance of top MONA models}
\begin{table}[ht]\label{tab:mona-val-test}
\caption{Validation and Test WER of crossCon+DTW 256k models.}
\vskip 0.1in
\centering
\begin{small}
\begin{sc}
\begin{tabular}{lcc}
\toprule
\textbf{Run ID} & \textbf{Validation (\%)} & \textbf{Test (\%)} \\
\midrule
GAD-984 & 20.63 & 22.17 \\
GAD-992 & 20.79 & 21.69 \\
GAD-986 & 21.26 & 21.75 \\
GAD-996 & 21.32 & 21.87 \\
GAD-993 & 21.45 & 20.72 \\
GAD-987 & 21.45 & 20.90 \\
GAD-988 & 21.63 & 20.96 \\
GAD-995 & 21.69 & 22.54 \\
GAD-983 & 21.82 & 22.11 \\
GAD-994 & 22.27 & 21.87 \\
\midrule
\textbf{Average} & 21.43 & 21.66 \\
\bottomrule
\end{tabular}
\end{sc}
\end{small}
\label{tab:crossCon_DTW_256k_models}
\end{table}

To explore to what extent validation performance is predictive of test performance, we conducted a Spearman rank correlation test on Table \ref{tab:crossCon_DTW_256k_models}. Spearman's rho was 0.22 (p-val: 0.56) indicating a weekly positive but not statistically significant relationship between validation WER and test WER when choosing between different training runs of the same model architecture differing only by initial seed.

\end{document}